\documentstyle[amsmath,amssymb,multicol,epsfig,prb,aps]{revtex}
\newcommand{\bra}{\langle}
\newcommand{\ket}{\rangle}
\newcommand{\approxlt}{\lesssim}
\newcommand{\approxbt}{\gtrsim}

\title{Melt viscosities of lattice polymers using a Kramers
  potential treatment.}
\author{O. D\"{u}rr${}^1$, H.L. Frisch${}^{2}$ and W. Dieterich${}^1$} 
\address{${}^1$Fachbereich Physik,
Universit\"{a}t Konstanz, D-78457 Konstanz, Germany\\
${}^2$Dept. of Chemistry, SUNY at Albany, Albany NY 12222, USA}
\begin{document}
\draft
\maketitle
\begin{abstract}
  Kramers relaxation times $\tau_{K}$ and relaxation times $\tau_{R}$
  and $\tau_{G}$  for the end-to-end distances and for center of mass
  diffusion are calculated for dense systems of athermal lattice
  chains. $\tau_{K}$ is defined from the response of the radius of
  gyration to a Kramers potential which approximately
  describes the effect of a stationary shear flow. It is shown that
  within an intermediate range of chain lengths $N$ 
  the relaxation times $\tau_{R}$ and $\tau_{K}$
  exhibit the same scaling with $N$, suggesting that
  $N$-dependent melt-viscosities for non-entangled chains can be
  obtained from the Kramers equilibrium concept.
\end{abstract}
\begin{multicols}{1}\narrowtext

\section{Introduction} 
Extending the novel approach by Kramers \cite{KRA46} for polymeric solutions
under shear (based on replacing the simple shear flow by its irrotational part)
\cite{FRI88,OHNXX,SCHXX} to melts 
allows us to compute a characteristic time $\tau_{K}$ related to the
melt viscosity.\cite{PAU99} In Ref.~5 the method was applied to alkane chains
up to about 50 carbon atoms. In this note we compare this concept for
melts with calculations of characteristic times for the end-to-end
relaxation of chains (rotational time $\tau_{R}$) and a time
$\tau_{G}$ obtained 
from the center of mass diffusion. This is achieved by Monte Carlo
simulations for dense systems of self--avoiding lattice chains on a
simple--cubic 
lattice. Chains up to a length $N=120$ are considered, which near the
upper limit already indicate the onset of entanglements. Our
calculations show that there exists an intermediate range of chain
lengths $N$ where $\tau_{K}$ and $\tau_{R}$ agree in
their dependence on $N$. 
This confirms the conclusion that important information on the melt
viscosity as a function of $N$ can be drawn from knowledge of fourth and
sixth moments of the equilibrium  distribution function for beads of the
chain. In the next section we describe our lattice model, followed in
section III by the description of our relaxation time results.

\section{Lattice--Model and Kramers relaxation time $\tau_{K}$}
We model the self--avoiding polymer by athermal lattice chains  where
every point of a simple cubic lattice with unit spacing is allowed to
be occupied by at most one chain monomer. 
In this way the hard-core repulsion is taken into account. 
For the dynamics and for the equilibration of
our system we use the generalized Verdier-algorithm, \cite{VER69}
consisting of local moves of one or two monomers. 
This allows us to determine the center of mass coordinates, the center
of mass diffusion constant and  the end
to end distance as a function of Monte-Carlo time steps, for a number
of chain densities. The average chain 
densities are measured by the number of monomers per lattice point.
We investigate monomer densities $n$ ranging from a single chain up
to $n=0.5$ (half filled lattice).

In Ref.~2 it was shown that a relaxation time $\tau_{K}$ can be
obtained from the Kramers potential \cite{KRA46} for Rouse chains in
a simple shear flow. This time is obtained from the
equilibrium averaged fourth and sixth moments of chain monomer coordinates
$x_{i}$ and $y_{i}$ relative to the center of mass. If $\bra X^{2}_{G} \ket_{o}$ 
is the mean squared x--component of the radius of gyration of the
chain at equilibrium, then \cite{fuss1}

\begin{multline}
  \tau_{K}^{2} = \frac{1}{2} \left[ \bra X^{2}_{G} \ket_{o}^{-1} \frac{1}{N}
  \sum_{i=1}^{N} \sum_{j=1}^{N} \sum_{k=1}^{N} \bra
  x_{i}^{2} x_{j} x_{k} y_{j} y_{k} \ket_{0} \right.
 - \\ \left. \sum_{j=1}^{N}\sum_{k=1}^{N} \bra x_{j}x_{k}y_{j}y_{k}\ket_{0} \right]
\label{eq:1}
\end{multline}


where  $x_{i}$ and $y_{i}$ are the coordinates of the chain
monomers with respect to the center of mass of a chain.
In the Rouse model one can calculate the shear viscosity $\eta$ from this
relaxation time using
\begin{equation}
  \eta = \frac{\pi^{2} \rho N_{A} k_{B} T \tau}{12 M}
  \label{eq:2}
\end{equation}
where $\rho$ is the mass density, $N_{A}$ is Avogadro's number, $M$
the molar mass of the molecule and $\tau$ either the Rouse or the
reduced Kramers time given by (\ref{eq:1}). 
With $\tau$ equal to the reduced Kramers time $\tau_{K}$, 
Eq.~\ref{eq:2} was shown
earlier\cite{PAU99} to give the 
melt (shear) viscosity of shorter alkanes up  to about 50
$\text{CH}_{2}$-- units. 
In this note we do not obtain $\eta$ directly but we calculate 
and compare $\tau_{K}$, $\tau_{R}$ and $\tau_{G}$ as a function of
chain length $N$ at various monomer densities.
While $\tau_{K}$ is calculated from (\ref{eq:1}), we obtain $\tau_{R}$
by fitting an exponential to the decay of the end-to-end vector
correlation function at large times. The time $\tau_{G} = \bra R^{2}_{G} \ket /D$
is obtained from simulated center of mass diffusion
coefficients D and the mean square radius of gyration $\bra R^{2}_{G}
\ket$.

\section{Simulation Results for $\tau_{K}$, $\tau_{R}$ and $\tau_{G}$}
As a background to our further results we show in Fig.~1a
plot of the diffusion constants $D$ divided by the Rouse diffusion
coefficient $D_{0}=(1-n)/N$ for various densities. Here the factor $1-n$
has been introduced to account for the average effect of blocking of
sites by other monomers.
The drop in $D/D_{o}$ for sufficiently large $N$, seen in
Fig.~1, can be thought to herald the onset of reptation. 
The horizontal line represents the
expected $N$-dependences of the free chain while the line
with slope $-1$ reflects the expected behavior for reptating chains. 
The inset provides an approximate master curve for the diffusion
constant versus $N/N_{e}$ where $N_{e}$ is an estimated length for the onset of
entanglements.\cite{PAU91}

Fig.~2 presents double--logarithmic plots of the various
relaxation times against $N$ for three different densities. Naturally,
for short chains one wouldn't expect any universal properties so that
differences in the behaviors of $\tau_{K}$, $\tau_{R}$ and $\tau_{G}$
are not surprising. In the dilute limit, displayed in Fig.~2a, both $\tau_{K}$ and $\tau_{R}$ grow with $N$
approximately as $(N-1)^{2.26}$ when $N \approxbt 20$, in reasonable
agreement with earlier numerical studies.\cite{FRI88,BIN95,SKO90} For the
purpose of illustrating this $N$--dependence common to $\tau_{K}$ and
$\tau_{R}$ we have multiplied $\tau_{K}$ by a constant factor so that
both sets of data fall on a common curve. The associated effective
exponent in this range of $N$--values is in fair agreement with the
theoretical expectation $\tau_{R} \sim N^{2\nu\!+\!1}$ (see Ref.~11) with
$\nu \approx 0.59$. For comparison, data for $\bra R^{2}_{G}\ket/N
\sim N^{2 \nu -1}$ are also shown in Fig.~2a. On the
other hand, with $D \sim N^{-1}$ for unentangled chains, it is
clear that $\tau_{G}$ will also be proportional to $N^{2\nu+1}$.

A different behavior is observed in Fig.~2b for a
monomer density $n=0.3$. 
Apart from a constant numerical factor, the relaxation times $\tau_{K}$
and $\tau_{R}$ fall together in the range $10 \approxlt N \approxlt
10^{2}$. Deviations beyond $N \simeq 10^{2}$ are expected to be
caused by entanglement effects which then are no longer negligible and are
not included in the Kramers method.
Note that for large $N$ sufficiently above the entanglement length one
expects $\tau_{R} \sim N^{3.4}$ and $\tau_{G} \sim N^{3}$.
The effective dynamic exponent read from the slope of the straight
line in Fig.~2b in principle may depend on $n$ (see
Ref.~10) but, similar to Fig.~2a, is still found in
fair agreement with $2 \nu + 1$. On the other hand, screening effects now diminish the growth of $\bra R_{G}^{2}\ket$ with $N$ relative to the
case of a free chain shown in Fig.~2a. 
The stronger increase of $\tau_{G}$ with $N$ becomes clear from the
behavior of $D$, see Fig.~1.

These trends with $n$
are continued in Fig.~2c where the range of agreement between
$\tau_{K}$ and  $\tau_{R}$ shrinks to about $10 \approxlt N
\approxlt 30$. 
Fig.~2c also suggests that for $N$ substantially larger than
$N_{e}$ $\tau_{K}$, assumes again Rouse-like scaling behavior, i.e.  
$\tau_{K}\sim N^{2}$.

In passing, we note that for the ratio of the square of the end-to-end
distance $R$  to the squared radius of gyration $R_{G}$ 
saturates with $N$ and comes close to the value 6 for our higher
densities, as expected for Gaussian chains, whereas for lower
densities this ratio is 
significantly larger than 6. This ratio versus $N$ is plotted in Fig.~3. 
\section{Conclusion}
The Kramers potential approach for the viscosity relaxation time of
dense systems of lattice chains has been verified numerically for an
intermediate range of chain lengths. For longer chains the breakdown of
this concept could be identified as a result of the onset of
entanglement effects. Generally, the Kramers approach offers the
possibility to obtain information on a dynamic quantity from
equilibrium data. Equilibrium lattice chain configurations can be
generated very efficiently via the configurational bias
method.\cite{SIE92} From such data the static moments in Eq.~1
and hence information about melt viscosities in a certain range of
densities and chain lengths can be obtained with minimal computational effort.
\section{Acknowledgments}
We acknowledge helpful discussion with
F.\ Eurich and P.\ Maass. H.\ L.\ Frisch was supported in part also by the NSF grant
DMR 9623224, the Humboldt Foundation and Donors of the Petroleum Fund of the
American Chemical Society.

\newpage
\section*{Figures}
Fig.~1. Chain-length dependence of the diffusion coefficient for
        a free chain ($\square$), $n=0.1$ $(\ast)$, $n=0.3$ $(\times)$ and
        n=0.5 (+). By using $D^{*}_{0}=k^{*}/N$ with $k^{*}$ being
        chosen as a free parameter for each $n$ and by rescaling the N
        axis via $N \rightarrow N/N_{e}$ all curves (except the one
        for free chains) fall together on one curve, as shown in the inset. The
        values chosen for $N_{e}$ are 23 for $n=0.5$, 50 for $n=0.3$ and
        150 for $n=0.1$ (and $k^{*}=0.50,0.72,1.28$ respectively).

\vspace{1cm}
Fig. 2.  Comparison of the reduced Kramers relaxation times
  $\tau_{K}/(N\!-\!1)^{2}$ $\times$  with the relaxation time of the end-to-end vector
  $\tau_{R}/(N\!-\!1)^{2}$ ($\square$) and $\tau_{G}/(N\!-\!1)^{2}$ ($\ast$) for
  different number densities $n$, a) free chains,
  b) $n=0.3$ and c) $n=0.5$. Also included is the
  radius of gyration squared ($+$) divided by $N-1$. Typical error
  bars for larger $N-$values are indicated in Fig.~2c,
  whereas the size of error bars in the intermediate range, where
  $\tau_{R}$ and $\tau_{K}$ coincide, is comparable to or smaller than
  the symbol size.
\vspace{1cm}

Fig.~3. Comparison of $R^{2}/R_{G}^{2}$ vs. chain length $N$
        for different concentrations. Also included is the
        $N \rightarrow \infty$ limit for Gaussian chains.

\begin{figure}[htbp]
  \begin{center}
     \epsfig{file=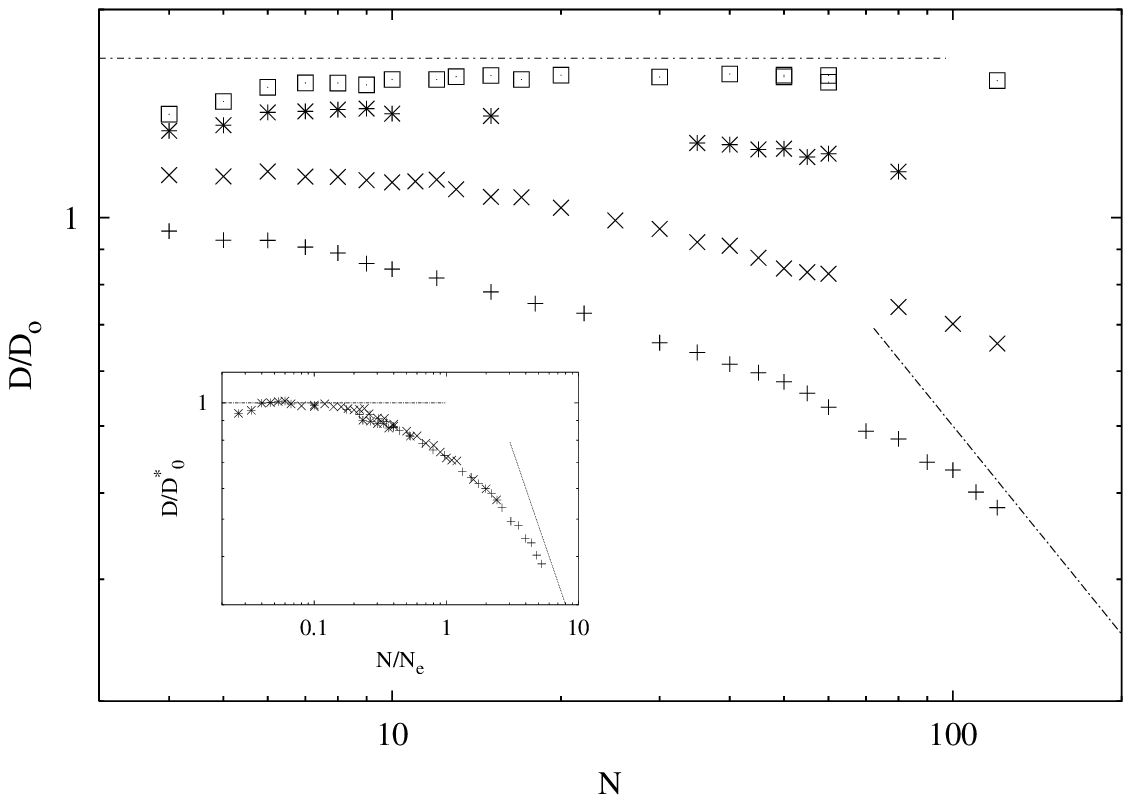,width=1.0\linewidth} 
     \label{fig:1}
     \section*{Fig. 1.}
  \end{center}
\end{figure} 

\begin{figure}[htbp]
\begin{center}
\epsfig{file=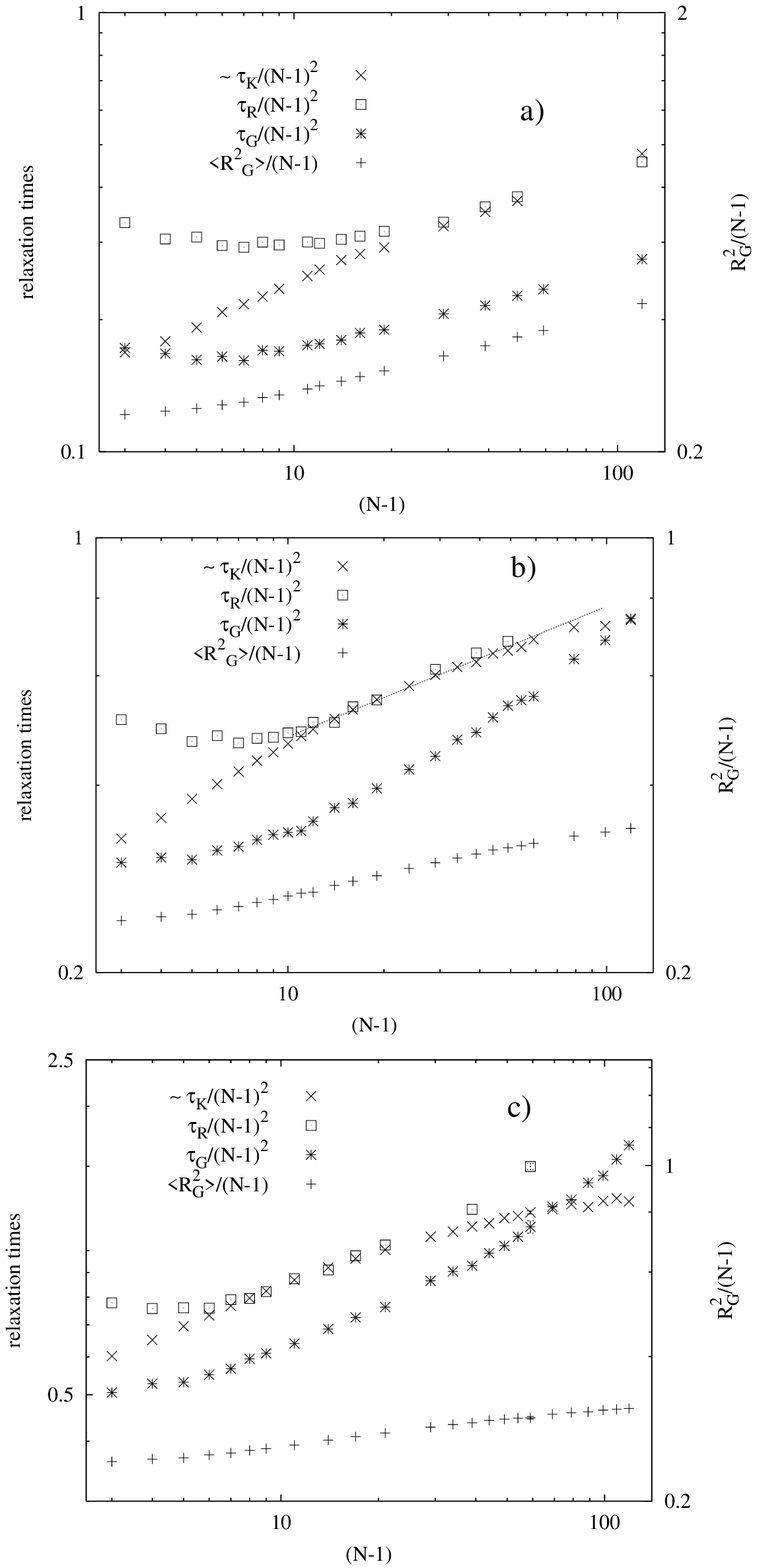,width=1.0\linewidth}
\section*{Fig. 2.}
\label{fig:2}
\end{center}
\end{figure}

\pagebreak
\begin{figure}[htbp]
  \begin{center}
     \epsfig{file=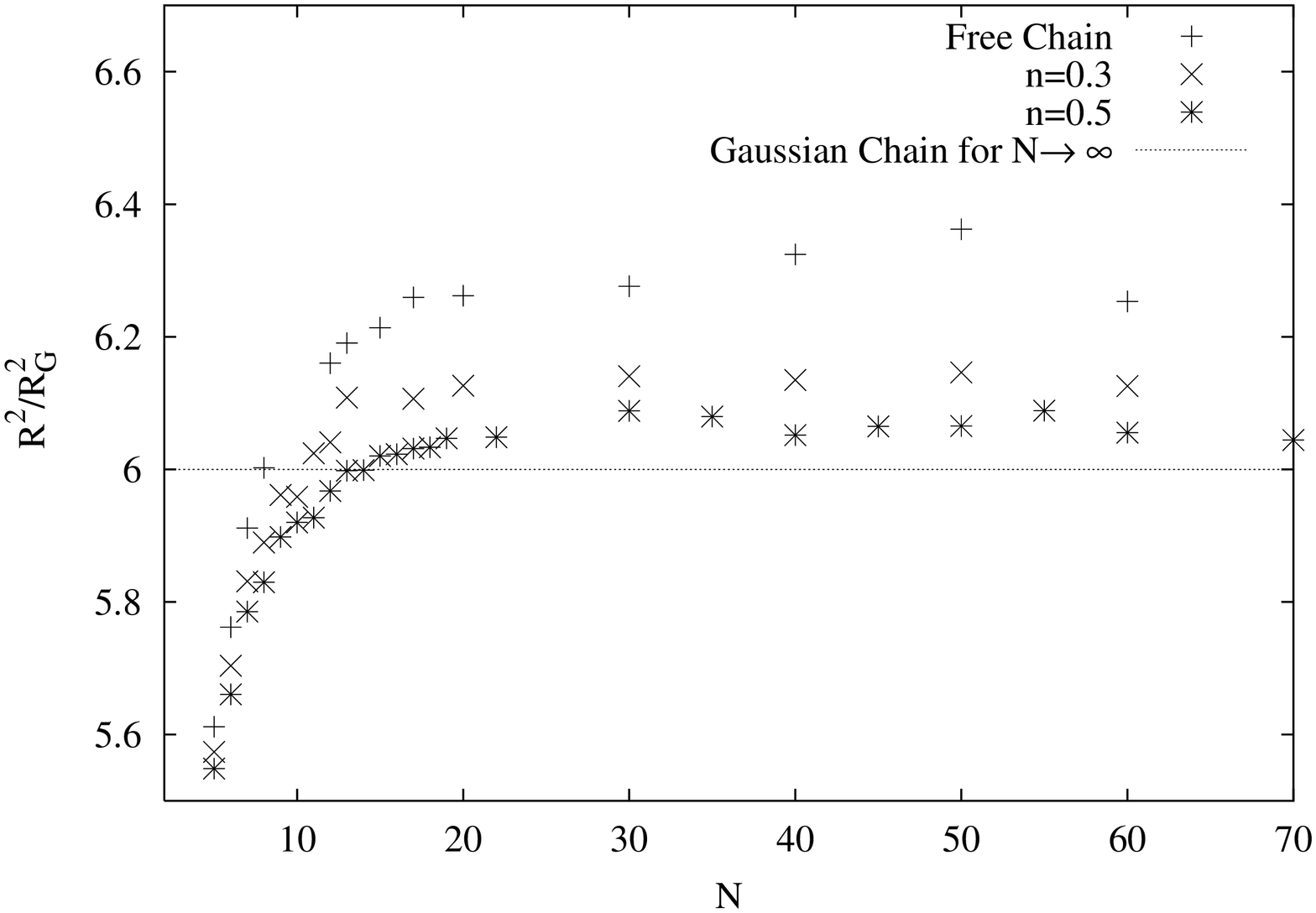,width=0.60\linewidth,angle=-90} 
     \label{fig:3}
     \section*{Fig. 3.}
  \end{center}
\end{figure} 
\newpage
\end{multicols}

\begin{thebibliography}{99}
\bibitem{fuss1} Up to a numerical factor, the time unit is
  $\zeta/k_{B}T$, where $\zeta$ is the friction coefficient in the
  equivalent Rouse model.\cite{FRI88}
\end{thebibliography}
\end{document}